\documentclass[aps,pra,reprint,superscriptaddress,twocolumn, longbibliography]{revtex4-2}

\pdfoutput=1

\usepackage{graphicx}
\usepackage{amsmath,amsfonts,amssymb,mathtools}
\usepackage{color}
\usepackage{subfigure}
\usepackage{physics}
\usepackage{dsfont}
\usepackage{hyperref}
\usepackage{booktabs}
\usepackage{url}
\usepackage{leftidx}
\usepackage{textcomp}
\usepackage{gensymb}
\usepackage{threeparttable}
\usepackage{rotating}
\usepackage{esvect}
\usepackage{booktabs}
\usepackage{marvosym}
\usepackage{lmodern}
\usepackage{mhchem}
\usepackage[noend]{algpseudocode}
\usepackage[dvipsnames]{xcolor}
\usepackage{multibib}
\newcites{supp}{Supplementary}

\definecolor{emerald}{rgb}{0.31, 0.78, 0.47}
\definecolor{blue(ncs)}{rgb}{0.0, 0.53, 0.74}

\hypersetup{
     colorlinks=true,
     linkcolor=magenta,
     filecolor=blue,
     citecolor=emerald,      
     urlcolor =blue(ncs),
}

\usepackage{chngcntr}

\DeclareMathAlphabet{\pazocal}{OMS}{zplm}{m}{n}

\newcommand{\bo}[1]{\boldsymbol{#1}}

\newcommand{\D}{\mathrm{d}}
\newcommand{\f}[2]{\frac{#1}{#2}}

\newcommand{\AVS}{$A$V$_3$Sb$_5$}

\newcommand{\CVS}{CsV$_3$Sb$_5$}
\begin{document}

\title{Superconductivity from Orbital-Selective Electron-Phonon Coupling in \AVS{}}

\author{Ethan T.~Ritz}
\thanks{These authors contributed equally to this work.}
\affiliation{Department of Chemical Engineering and Materials Science, University of Minnesota, MN 55455, USA}

\author{Henrik S.~R{\o}ising}
\thanks{These authors contributed equally to this work.}
\affiliation{Niels Bohr Institute, University of Copenhagen, DK-2200 Copenhagen, Denmark}

\author{Morten H.~Christensen}
\affiliation{Niels Bohr Institute, University of Copenhagen, DK-2200 Copenhagen, Denmark}

\author{Turan Birol}
\affiliation{Department of Chemical Engineering and Materials Science, University of Minnesota, MN 55455, USA}

\author{Brian M.~Andersen}
\affiliation{Niels Bohr Institute, University of Copenhagen, DK-2200 Copenhagen, Denmark} 

\author{Rafael M.~Fernandes}
\affiliation{School of Physics and Astronomy, University of Minnesota, Minneapolis,
MN 55455, USA}
\date{\today}

\begin{abstract}
Recent experiments have shown that the phase diagrams of the kagome superconductors \AVS{} are strongly impacted by changes in the $c$-axis lattice parameter. Here, we show that $c$-axis deformations impact primarily the Sb apical bonds and thus the overlap between their $p_z$ orbitals. Changes in the latter, in turn, substantially affect low-energy electronic states with significant Sb character, most notably the central electron pocket and the van Hove singularities located above the Fermi level. Based on the orbital-selective character of $c$-axis strain, we argue that these electronic states experience a non-negligible attractive electron-phonon pairing interaction mediated by fluctuations in the apical Sb bonds. We thus propose a multi-band model for superconductivity in \AVS{} that includes both the Sb pocket and the V-derived van Hove singularities. Upon comparing the theoretical phase diagram with the experimentally observed vanishing of the $T_c$ dome across a Lifshitz transition of the Sb pocket, we propose that either an $s^{+-}$ or an $s^{++}$ state is realized in \AVS.
\end{abstract}

\maketitle

The discovery of superconductivity (SC) in the family of kagome metals \AVS{} ($A$: K, Rb, Cs) has sparked significant interest, since the interference between different electronic hopping paths in the kagome lattice endows the electronic structure with flat bands, van Hove singularities (vHs), and Dirac points. These features have the potential to promote collective electronic behaviors characteristic of materials with strong electronic correlations or non-trivial band topology~\cite{Yin2018Giant,Ghimire2020Topology, Kang2020Dirac, Gilmutdinov2021Interplay}. Indeed, upon a cursory examination, the phase diagrams of \AVS{} resemble those of Cu- and Fe-based superconductors, in that SC appears in close proximity to another electronic order, in this case a charge-density wave (CDW) phase, which has been intensely scrutinized both theoretically~\cite{ParkEA21, LinEA21, Denner2021,Tazai2022mechanism, Christensen2021, Ferrari2022,Christensen2022} and experimentally~\cite{Ortiz2019New, Kenney2021Absence, Jiang2021Unconventional, Chen2021Roton, Zhao2021Cascade}. While the three-dimensional nature of the CDW wave-vector is well established \cite{RatcliffEA21, Stahl2022Temperature-driven, Wu2022Charge, Kang2022Charge}, there remains considerable debate  whether it also breaks time-reversal and rotational symmetries~\cite{Yang2020Giant, Xiang2021Twofold, Mielke2022Time-reversal, Li2022Rotation, Xu2022Three-state, Nie2022Charge-density-wave-driven, Guo2022Switchable, Saykin2022High}.

Studies of the SC properties of \AVS{} have lagged the investigations of the CDW phase, partly because the latter onsets at much higher temperatures ($T_{\mathrm{CDW}} \sim 100$ K) than the former ($T_c \sim 1$ K). There have been reports of both nodeless~\cite{Duan2021Nodeless, Gupta2022Microscopic, Gupta2022Two, RoppongiEA22, Mu2021S-wave,Xu2021Multiband,Zhang2023} and nodal~\cite{Zhao2021nodal,Chen2021Roton,Guguchia2022Tunable} gap structures, as well as conflicting accounts of whether the electron-phonon coupling can explain the SC instability~\cite{TanEA21,Zhang2021firstprinciples,Zhong2022Testing,Wang_phonon_2023}. Proposals have been put forward in favor of both conventional and unconventional pairing, most of which focus on the electronic states derived from the V $d$-orbitals~\cite{ParkEA21,LinEA21,WeEA21,He2022Strong-coupling,Tazai2022mechanism,Wen2022superconducting,Bai2022effective,RomerEA22}, which give rise to saddle points in the band structure near the M point. However, recent experimental studies of the doping-temperature and pressure-temperature phase diagrams of \CVS{} have shown that the end of the SC dome in either case coincides with the disappearance of an electron pocket at the $\Gamma$ point derived from the Sb $p$-orbitals, highlighting their relevance for the onset of pairing~\cite{TsirlinEA22,OeyEA22}.

In this paper, we combine density functional theory (DFT) calculations and low-energy modeling to show that the Sb degrees of freedom play an essential role for the superconductivity of \AVS{}. Our starting point is the empirical observation made in Ref.~\onlinecite{Qian2021Revealing} that the phase diagrams of \CVS{} under pressure and under uniaxial in-plane stress fall essentially on top of each other when $T_{\mathrm{CDW}}$ and $T_c$ are expressed as a function of the $c$-axis expansion or contraction. Moreover, thermal expansion measurements report a $c$-axis response much more pronounced than the $a$-axis response at both $T_c$ and $T_{\mathrm{CDW}}$~\cite{MeingastAPS}. These results are strong indication that the $c$-axis lattice parameter is a key control parameter for both types of electronic order. From DFT, we find that the states primarily affected by changes in the $c$-axis are those that have a significant contribution from the apical Sb orbitals: the vHs above the Fermi level at the M point and the electron band centered at the $\Gamma$ point, whose bottom shifts by hundreds of meV for strain values of a few percent. In contrast, the energies of the vHs below the Fermi level remain nearly unchanged. Such an ``orbital-selective" modification of the electronic spectrum provides a mechanism by which the $c$-axis lattice parameter can impact the CDW and SC transitions, as empirically seen in Ref.~\onlinecite{Qian2021Revealing}. Based on these results, we construct a low-energy model for the SC state of \AVS{} consisting of a central Sb-dominated electron pocket and a large V-dominated Fermi surface associated with the vHs. By studying the evolution of $T_c$ as the Sb band raises above the Fermi level, we find that only $s^{++}$ and $s^{+-}$ states are compatible with the observation of a vanishing $T_c$ as the Sb pocket undergoes a Lifshitz transition, in agreement with experiments~\cite{TsirlinEA22,OeyEA22}.

\begin{figure*}
    \centering
    \includegraphics[width=\textwidth]{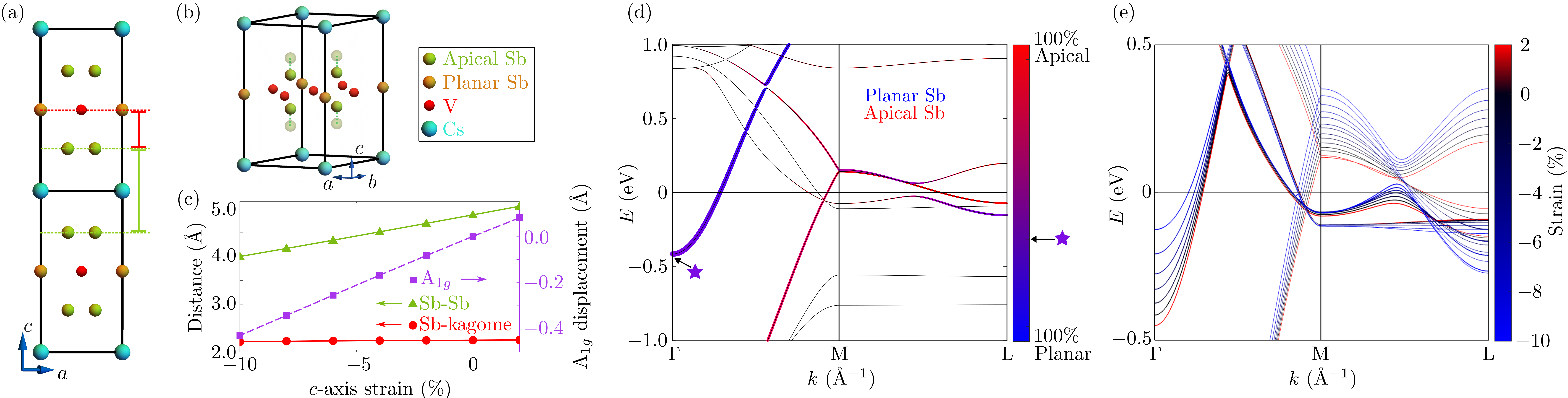}
    \caption{\label{fig:dft_results} (a)-(b) Schematic illustrations of two neighboring \AVS{} unit cells and of the displacement pattern of the apical Sb promoted by the A$_{1g}$ phonon mode, respectively. We choose a sign convention such that the displacement of apical Sb towards the Kagome plane corresponds to positive A$_{1g}$. (c) Changes in the distances between the Sb apical atom and its nearest-neighbor (green triangles) and the V atoms (red circles) as a function of $c$-axis strain for \CVS{}. The purple squares give the atomic displacements from the equilibrium structure corresponding to a frozen excitation of the optical A$_{1g}$ phonon mode. (d) Low-energy band structure of \CVS{}. The thickness of the bands is proportional to the total projection onto the $p_z$ orbitals of the Sb atoms, whereas their color is proportional to the projection onto the planar Sb (blue) and apical Sb (red). The $\Gamma$ band has contributions from both Sb sites. (e) Modifications in the low-energy band structure as a function of $c$-axis strain. The momentum L is located above M in the hexagonal Brillouin zone; the other momenta are defined in Fig.~\ref{fig:PatchModel}(a).}
\end{figure*}

We start by employing DFT to elucidate the impact of $c$-axis distortions on the band structure of \AVS{}, focusing on $A=\mathrm{Cs}$ for concreteness. For details, see the Supplementary Material (SM)~\cite{SM}. Above $T_{\mathrm{CDW}}$ and at ambient pressure, \CVS{} adopts the $P6/mmm$ (\#191) space group, with Cs occupying the $1a$ Wyckoff site, V the $3g$ site, and Sb the $4h$ (apical) and $1b$ (planar) sites, as illustrated in Figs.~\ref{fig:dft_results}(a)-(b). The V atoms form a kagome sublattice, whereas the planar  (apical) Sb atoms, a hexagonal (honeycomb) sublattice. Besides the lattice parameters, the reduced $z$ coordinate of the apical Sb atoms is the only free structural parameter. Interestingly, $z$ increases significantly upon compression of the $c$-axis, in a way that approximately preserves the Sb--V bond distances between the apical Sb and the kagome layer, while shortening the Sb--Sb bond distances between apical Sb in adjacent unit cells, as shown Fig.~\ref{fig:dft_results}(c). \phantom{\cite{kresse1993,kresse1996efficient,kresse1996efficiency,BeugelingEA12,GuoEA09,GuEA22,Sauls94}}

To address whether this displacement pattern is capable of affecting the low-energy electronic states, we first calculate via DFT the atomically-resolved band structure near the Fermi energy in the undistorted phase [Fig.~\ref{fig:dft_results}(d)]. In agreement with previous works  \cite{TsirlinEA22,OeyEA22,JeongEA22}, we find dominant spectral-weight contributions from both types of Sb atoms (planar and apical) to the $\Gamma$-point electron band, as well as a significant contribution from the apical Sb to the V-dominated saddle points located above the Fermi level at the M point. It is this hybridization between apical Sb orbitals and V orbitals that endow the corresponding vHs with a significant $k_z$ dispersion, to the point that they even cross the Fermi level along the M--L line. 

In Fig.~\ref{fig:dft_results}(e), we show how the low-energy band structure is modified by both compressive (negative) and tensile (positive) $c$-axis strain (see also Ref.~\onlinecite{Consiglio2022}). We include large absolute values of strain to highlight the effect. Note that all internal lattice parameters are relaxed for a given $c$-axis distortion, while keeping the in-plane lattice parameter fixed. The bands that are most affected are those exhibiting a sizable contribution from the apical Sb atoms, such as the vHs located above the Fermi level. Since the CDW is associated with the condensation of phonon modes at the M and L points~\cite{RatcliffEA21,Christensen2021}, this provides a possible mechanism by which $c$-axis strain can impact the CDW phase. Besides these saddle points, the bottom of the electron pocket at the $\Gamma$ point moves substantially with $c$-axis changes, with shifts of $100$ meV for strains of about $1\%$ (see also Fig. S1 in the SM). In contrast, the energies of the M-point vHs located below the Fermi level barely change, reflecting their dominant V character. The large shifts of the bottom of the $\Gamma$-point electron band can be attributed to the out-of-phase overlap between the $p_z$ orbitals of apical Sb atoms of neighboring unit cells. This overlap generates a bonding and an anti-bonding state, the latter of which gives rise to the $\Gamma$-point electron band. Upon compression of the $c$-axis, the orbital overlap increases and, consequently, the energy of the anti-bonding state increases, leading to the observed shift in the bottom of the electron band.

The electronic properties of \AVS{} should be impacted not only by static strain, but also by thermal fluctuations associated with the atomic displacement pattern promoted by the $c$-axis strain. These fluctuations are expected to be strongly coupled to the electronic states with significant Sb character. Because the displacement pattern associated with the Sb--Sb bonds does not break crystal symmetries, it cannot be decomposed in terms of a single phonon mode. Instead, there are two different phonon modes that modify the bond lengths along the $c$-axis without modifying other features in the crystal structure: a longitudinal acoustic phonon mode with out-of-plane dispersion and a $\Gamma$-point optical phonon mode that transforms as the $A_{1g}$ irreducible representation of the point group. Note that the $A_{1g}$ displacements, represented in Fig.~\ref{fig:dft_results}(b), also involve changes in the Sb--V bond distances. As shown in Fig.~\ref{fig:dft_results}(c), the strong $c$-axis strain dependence of the displacement associated with this $A_{1g}$ mode resembles that displayed by the Sb--Sb bond distance. This result confirms that the $A_{1g}$ phonon mode leads to Sb--Sb bond fluctuations.

\begin{figure*}
	\centering
	\includegraphics[width=\linewidth]{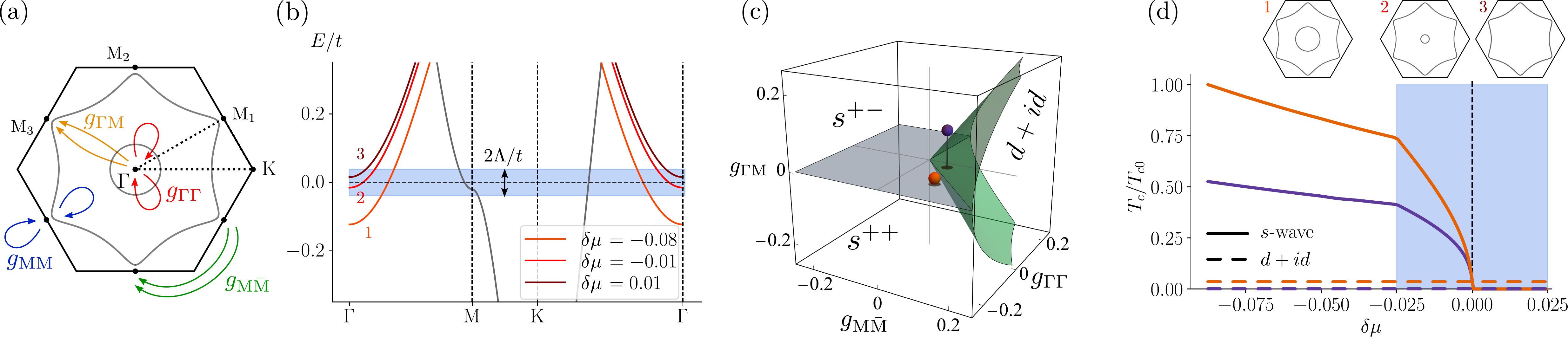}
    \caption{(a) Pairing interactions of the four-patch model involving fermions at the M$_i$ and $\Gamma$ points. The Fermi surface is shown in gray. (b) Tight-binding dispersions, highlighting the Lifshitz transition of the $\Gamma$-point electron pocket as a function of the parameter $\delta \mu \equiv (\mu_{\Gamma}-\mu_c)/\mu_c$; $\Lambda$ is the pairing interaction cutoff. (c) SC phase diagram of the four-patch model (away from the Lifshitz transition) as a function of the interactions shown in panel (a);
    $s$-wave corresponds to either $s^{++}$ or $s^{+-}$ states depending on the sign of $g_{\Gamma \mathrm{M}}$. (d) $T_c$ as a function of the parameter $\delta \mu$ that tunes the $\Gamma$-pocket across the Lifshitz transition at $\delta \mu =0$,  as shown in the insets. The interaction parameters, marked by the orange and purple symbols shown in (c), are ($g_{\rm M \bar{M}}, g_{\Gamma \mathrm{M}}, g_{\Gamma \Gamma},g_{\rm MM}$)=($0.1,0.015,-0.15,0$) for the orange lines and ($0.07,0.1,-0.03,0$) for the purple lines. $T_{c0}$ is the SC transition temperature for the first set of parameters (orange) at $\mu_{\Gamma 0} = -3.65 \tilde{t}$.} 
	\label{fig:PatchModel}
\end{figure*}

The coupling between the electronic states with significant Sb spectral weight and these phonon modes should lead to a non-negligible attractive pairing interaction. To assess its impact, we construct a low-energy model for the SC phase considering a simplified Fermi surface that consists of a small Sb electron-pocket at the $\Gamma$ point and a large hexagonal-like Fermi surface originating from one of the V vHs~\cite{WeEA21, RomerEA22, JeongEA22, TsirlinEA22}. The V band is modeled in terms of a single orbital on the sites of the kagome lattice whereas the Sb band is parametrized as a nearly isotropic dispersion, $\xi_{\Gamma}(\bo{k}) = \tilde{t} f_{\Gamma}(\bo{k}) - \mu_{\Gamma}$; details are given in the SM~\cite{SM}. The parameter $\mu_\Gamma$ defines the energy of the bottom of the electron band, and is set to $\mu_{\Gamma 0} = -3.65 \tilde{t}$ for the undistorted compound based on comparison with ARPES measurements~\cite{KangEA22}. Upon decreasing $\mu_\Gamma$, which mimics the effect of hole doping, a Lifshitz transition occurs at $\mu_c \equiv -4 \tilde{t}$, where the $\Gamma$-point Fermi pocket disappears. For simplicity, we thus define $\delta \mu \equiv (\mu_{\Gamma}-\mu_c)/\mu_c$.

To derive the SC gap equations, we generalize a patch approach commonly employed to describe systems with vHs near the Fermi level ~\cite{NandkishoreEA12,IsobeEA18,CLassen2020,Chichinadze2020,Chichinadze2021,WuEA22}. Because of the logarithmic enhancement of the density of states (DOS) at the M-point vHs, it is sufficient to consider only the pairing interactions involving states on the three Fermi surface patches centered at each of the three M points. Symmetry restricts these interactions to two different types: intra-patch $g_{\mathrm{MM}}/N_{\mathrm{M}}$ and inter-patch $g_{\mathrm{M}\bar{\mathrm{M}}}/N_{\mathrm{M}}$, where $N_{\mathrm{M}}$ is the DOS of the M-point patches. We approximate the small $\Gamma$ pocket by a fourth patch subjected to an intra-patch pairing interaction $g_{\Gamma \Gamma}/N_\Gamma$ and an inter-patch interaction $g_{\Gamma \mathrm{M}}/\sqrt{N_{\mathrm{M}} N_\Gamma}$ with the M-point patches. Based on our results above, we assume an attractive interaction $g_{\Gamma \Gamma}<0$ arising from the electron-phonon coupling involving the apical Sb degrees of freedom. Note that this parametrization of the pairing interaction in terms of the DOS of each patch is not valid close to the Lifshitz transition; we will return to this point later. 

The resulting four-patch model is schematically shown in Fig.~\ref{fig:PatchModel}(a). Denoting $\vec{\Delta} \equiv (\Delta_{\rm M_1}, \Delta_{\rm M_2}, \Delta_{\rm M_3}, \Delta_{\Gamma}
)^{\mathrm{T}}$ for the gap functions on the four patches, the corresponding linearized gap equations can be written in matrix form as $\chi_{\mathrm{pp}} \vec{\Delta} = \vec{\Delta}$, with
\begin{equation}
\chi_{\mathrm{pp}} = - V_{\Lambda}\begin{bmatrix}
 g_{\mathrm{MM}} & g_{\mathrm{M} \bar{\mathrm{M}}} & g_{\mathrm{M} \bar{\mathrm{M}}} & \eta g_{\Gamma \mathrm{M}} \\
g_{\mathrm{M} \bar{\mathrm{M}}} &  g_{\mathrm{MM}} & g_{\mathrm{M} \bar{\mathrm{M}}} &  \eta g_{\Gamma \mathrm{M}} \\
g_{\mathrm{M} \bar{\mathrm{M}}} & g_{\mathrm{M} \bar{\mathrm{M}}}&  g_{\mathrm{MM}} & \eta g_{\Gamma \mathrm{M}} \\
\eta^{-1} g_{\Gamma \mathrm{M}} &   \eta^{-1} g_{\Gamma \mathrm{M}} &  \eta^{-1} g_{\Gamma \mathrm{M}} &  g_{\Gamma \Gamma} 
\end{bmatrix}, \label{eq:Gmatrix}
\end{equation}
where $V_{\Lambda} \equiv \int_{-\Lambda}^{\Lambda} \D\varepsilon~\tanh(\beta \varepsilon/2)/(2\varepsilon)\approx \ln\left(2e^{\gamma} \beta \Lambda /\pi \right)$ is the particle-particle bubble with $\beta = 1/(k_B T)$ and $\eta \equiv \sqrt{N_\Gamma/N_{\mathrm{M}}}$ is the ratio between the DOS. Here, $\Lambda$ is the cutoff for the pairing interaction, as shown in Fig.~\ref{fig:PatchModel}(b), and $\gamma \approx 0.577$ is Euler's constant. $T_c$ is found by imposing that the largest eigenvalue of $\chi_{\mathrm{pp}}$ is 1. The two leading eigenvalues are
\begin{align}
\lambda_{E_{2g}} &= \left(g_{\mathrm{M} \bar{\mathrm{M}}}-g_{\mathrm{MM}}\right) \ln\left(2e^{\gamma} \beta \Lambda /\pi \right), \label{eq:E2eig} \\
\lambda_{A_{1g}} &= \frac12 \left(\tilde{g}-2g_{\mathrm{M} \bar{\mathrm{M}}}-g_{\mathrm{MM}}-g_{\Gamma \Gamma} \right) \ln\left(2e^{\gamma} \beta \Lambda /\pi \right), \label{eq:A1eig}
\end{align}
where $\tilde{g} \equiv \sqrt{12g_{\Gamma \mathrm{M}}^2 + (g_{\Gamma \Gamma}  - 2g_{\mathrm{M} \bar{\mathrm{M}}} - g_{\mathrm{MM}})^2}$. Analysis of the eigenvectors of $\lambda_{E_{2g}}$, which is doubly-degenerate, shows that they describe $d_{x^2-y^2}$-wave and $d_{xy}$-wave SC states, which transform as the 2D irreducible representation $E_{2g}$ of the point group $D_{6h}$ (see SM~\cite{SM}). As discussed elsewhere~\cite{NandkishoreEA12}, going beyond the linearized gap equation reveals that the linear combination $d_{x^2-y^2} \pm i d_{xy}$ minimizes the free energy, 
leading to a time-reversal symmetry-breaking SC phase. The second eigenvalue $\lambda_{A_{1g}}$ corresponds to a pairing state that transforms as the trivial representation $A_{1g}$, corresponding to two isotropic gaps $\Delta_\Gamma$ and $\Delta_{\mathrm{M}}$. While the symmetry of this state is $s$-wave, there are two qualitatively different possible gap configurations  depending on the signs of $\Delta_\Gamma$ and $\Delta_{\mathrm{M}}$: an $s^{++}$ state, in the case of equal signs, or an $s^{+-}$ state, in the case of opposite signs -- similar to that realized in the Fe-based superconductors~\cite{Fernandes2022Iron}.

Only positive eigenvalues correspond to attractive pairing channels. For the $E_{2g}$ channel ($d+id$), $\lambda_{E_{2g}}>0$ requires a strong enough inter-M-patch repulsion to overcome the intra-M-patch repulsion, $g_{\mathrm{M} \bar{\mathrm{M}}} > g_{\mathrm{MM}} > 0$, as found in renormalization group studies of the three-patch model~\cite{NandkishoreEA12}. As for the $A_{1g}$ channel ($s^{++}$ or $s^{+-}$), $\lambda_{A_{1g}} > 0$ requires either a strong inter-patch interaction $g_{\Gamma \mathrm{M}}$, which can be repulsive or attractive, or a strong intra-$\Gamma$-patch attraction $g_{\Gamma \Gamma} < 0$. Note that the sign of $g_{\Gamma \mathrm{M}}$ does not impact the eigenvalue $\lambda_{A_{1g}}$, but only whether the eigenvector corresponds to the $s^{++}$ ($g_{\Gamma \mathrm{M}} < 0$) or the $s^{+-}$ ($g_{\Gamma \mathrm{M}} > 0$) state (see SM~\cite{SM}). 

Figure~\ref{fig:PatchModel}(c) shows the SC phase diagram in the $\{ g_{\mathrm{M} \bar{\mathrm{M}}},g_{\Gamma \Gamma},g_{\Gamma \mathrm{M}} \}$ parameter space. As anticipated, the $d+id$ state is only stabilized by a dominant repulsive interaction $g_{\mathrm{M} \bar{\mathrm{M}}}>0$, whereas attractive interactions of any kind favor an $s$-wave state. An increase in the magnitude of the inter-patch interaction $g_{\Gamma \mathrm{M}}$, be it attractive or repulsive, further expands the regime where the $s$-wave state is realized. While this plot is obtained for $g_{\mathrm{M} \mathrm{M}}=0$, the main effect of a non-zero $g_{\mathrm{M} \mathrm{M}}$ is in the case where it is repulsive, as it suppresses the regime in which SC is stabilized (see SM~\cite{SM}). 

To elucidate which of these SC regimes are consistent with the experimental observation of a suppression of $T_c$ across the Lifshitz transition~\cite{TsirlinEA22,OeyEA22}, we compute the evolution of $T_c$ as $\mu_{\Gamma}$ approaches the critical value $\mu_c$ for which the electron-band bottom crosses the Fermi level (see inset of Fig.~\ref{fig:PatchModel}(d)). Near the Lifshitz transition, where $\lvert \mu_{\Gamma} - \mu_c \rvert \ll \Lambda$, the gap equations~\eqref{eq:Gmatrix} have to be modified, as it is not justified to remove the DOS from the integrand of the particle-particle bubble~\cite{Fernandes2013,ChenEA15}. The modified $\chi_{\mathrm{pp}}$ is shown in the SM~\cite{SM}. By numerically computing its eigenvalues, we obtain $T_c(\mu_\Gamma)$ for the various regimes in Fig.~\ref{fig:PatchModel}(c). Because the $d+id$ state is insensitive to the $\Gamma$ pocket, its $T_c$ does not change across the Lifshitz transition. Meanwhile, the behavior of $T_c$ of the $s$-wave state depends on the nature of the dominant pairing interaction. If the $s$-wave state is driven by large attractive interactions involving the M patches only, $g_{\mathrm{M} \bar{\mathrm{M}}},\, g_{\mathrm{M} \mathrm{M}}<0$, $T_c$ is not significantly changed at $\mu_c$. On the other hand, for dominant intra-$\Gamma$-patch attraction $g_{\Gamma \Gamma}<0$ or dominant inter-patch $g_{\Gamma \mathrm{M}}$ of either sign, $T_c$ is strongly suppressed across the Lifshitz transition. This is shown in Fig.~\ref{fig:PatchModel}(d) for the parameter values corresponding to the orange symbol (dominant $g_{\Gamma \Gamma}$) and the purple symbol (dominant $g_{\Gamma \mathrm{M}}$) in Fig.~\ref{fig:PatchModel}(c). Additional $T_c(\mu_\Gamma)$ plots for other parameter values are shown in the SM~\cite{SM}. 

A large attractive intra-pocket pairing interaction $g_{\Gamma \Gamma}$ could be mediated by the Sb-Sb bond fluctuations discussed above. On the other hand, CDW fluctuations with wave-vector M could boost $g_{\Gamma \mathrm{M}}$, rendering it repulsive (attractive) if the CDW breaks (preserves) time-reversal symmetry. However, these CDW fluctuations should also enhance $g_{\mathrm{M} \bar{\mathrm{M}}}$, since the M patches are connected by the same wave-vector. Because the latter couples states with similar orbital compositions (V-V orbitals), whereas $g_{\Gamma \mathrm{M}}$ couples states with different orbital compositions (Sb-V orbitals), the CDW boost of $g_{\mathrm{M} \bar{\mathrm{M}}}$ is expected to be larger, particularly if the CDW is enhanced by the vHs. Interestingly, the Sb-Sb bond fluctuations could switch this hierarchy if the relevant vHs is one of those located above the Fermi level. Indeed, as shown in Figs.~\ref{fig:dft_results}(d)-(e), those vHs have a sizable Sb orbital weight, and as such should be impacted by the phonon modes associated with Sb-Sb bond displacements. 

We now discuss the experimental implications of our results. All three states obtained in our model, $d+id$, $s^{+-}$, and $s^{++}$, are fully gapped, which makes it challenging to distinguish between them solely via spectroscopy. Directly probing time-reversal symmetry breaking, for instance via Kerr rotation, would help exclude or confirm $d+id$. The suppression of $T_c$ across the Sb electron-pocket Lifshitz transition shown in Fig.~\ref{fig:PatchModel}(d) is qualitatively consistent with the experimental results of Refs.~\onlinecite{TsirlinEA22,OeyEA22}, suggesting that either an $s^{+-}$ or an $s^{++}$ state is realized, at least in the region of the phase diagram where the CDW is absent. These $s$-wave states are also compatible with the robustness of $T_c$ against impurities reported in Ref.~\onlinecite{RoppongiEA22} for \CVS{} and with the observed multi-gap structure of the SC state seen in the parent compounds, where SC coexists with CDW. If one of these gaps is small, it may reconcile reports favoring both a nodeless and a nodal pairing state~\cite{Duan2021Nodeless, Gupta2022Microscopic, Gupta2022Two, RoppongiEA22, Mu2021S-wave, Xu2021Multiband, Guguchia2022Tunable}. Alternatively, coexistence of an $A_{1g}$ SC state with CDW may lead to nodes in the reconstructed Fermi surface~\cite{Maiti2012}. As for unconventional SC, even if the $d+id$ state is subleading with respect to the $s^{+-}$ or $s^{++}$ channels, interesting mixed states can emerge when the ground states are close in energy. These include not only an $s+d+d$ state that has two-fold anisotropy, but also an $s+e^{i\theta}(d+id)$ state that breaks time-reversal symmetry, as discussed in Refs.~\onlinecite{Chichinadze2020,Wang_Fernandes2021}. Due to the presence of inversion symmetry, mixed singlet-triplet states are not expected.

In summary, we showed that changes in the $c$-axis lattice parameter of \AVS{} lead to significant changes in the electronic dispersion promoted by the apical Sb $p_z$ orbitals. Not only the energies and the $k_z$-dispersion of the vHs located above the Fermi energy are modified, but also the bottom of the $\Gamma$-point electron band shifts strongly with $c$-axis strain. We proposed that fluctuations of the Sb--Sb bonds promote a non-negligible electron-phonon pairing interaction for states with sizable Sb orbital character, which includes both the central electron pocket as well as the saddle points located above the Fermi level. The resulting $s^{+-}$ and $s^{++}$ states are consistent with several experimental observations, including the full suppression of $T_c$ across the Lifshitz transition involving the Sb pocket~\cite{TsirlinEA22,OeyEA22}, the nodeless gaps recently observed in ARPES~\cite{ZhongEA23}, as well as the robustness of $T_c$ against disorder~\cite{RoppongiEA22,Zhang2023}.

We thank N. Ni, Z. Wang, and S. Wilson for fruitful discussions. ETR and TB were supported by the NSF CAREER grant DMR-2046020. HSR was supported by research Grant No.~40509 from VILLUM FONDEN. MHC has received funding from the European Union's Horizon 2020 research and innovation programme under the Marie Sk{\l}odowska-Curie grant agreement No.~101024210. RMF was supported by the Air Force Office of Scientific Research under Award No. FA9550-21-1-0423.

\bibliography{Kagome}

\newpage
\widetext

\begin{center}
\textbf{\large --- Supplementary Material ---}
\end{center}

\setcounter{equation}{0}
\setcounter{figure}{0}
\setcounter{table}{0}
\makeatletter
\renewcommand{\theequation}{S\arabic{equation}}
\renewcommand{\thefigure}{S\arabic{figure}}

%
%%
%%%
\section{Details of the Density functional theory calculations}
\label{sec:DFT}
%%%
%%
%
All density functional theory (DFT) calculations were performed with Projector Augmented Waves (PAW) as implemented in the Vienna Ab initio simulation package (VASP) version 5.4.4 \cite{kresse1993,kresse1996efficiency,kresse1996efficient} using the PBEsol exchange-correlation functional
for valence configurations of Cs, V, and Sb corresponding to 5\emph{s}$^2$5\emph{p}$^6$6\emph{s}$^1$, 3\emph{s}$^2$3\emph{p}$^6$3\emph{d}$^4$4\emph{s}$^1$, and 5\emph{s}$^2$5\emph{p}$^3$, respectively. Lattice parameters were converged to within 0.001 {\AA} using a plane wave cutoff energy of 450 eV, combined with a $\Gamma$-centered Monkhorst-Pack k-point mesh of 20$\times$20$\times$10, as well as a Gaussian smearing parameter of 10 meV. Structural relaxation predicts equilibrium lattice parameters of $a=5.424$ {\AA} and $c=9.368$ {\AA}, and a reduced $z$ coordinate corresponding to apical Sb at the $4h$ $(1/3,2/3,z)$ Wyckoff site of $0.740$ in fractional coordinates.

\begin{figure}[t!bh]
\includegraphics[width=0.35\textwidth]{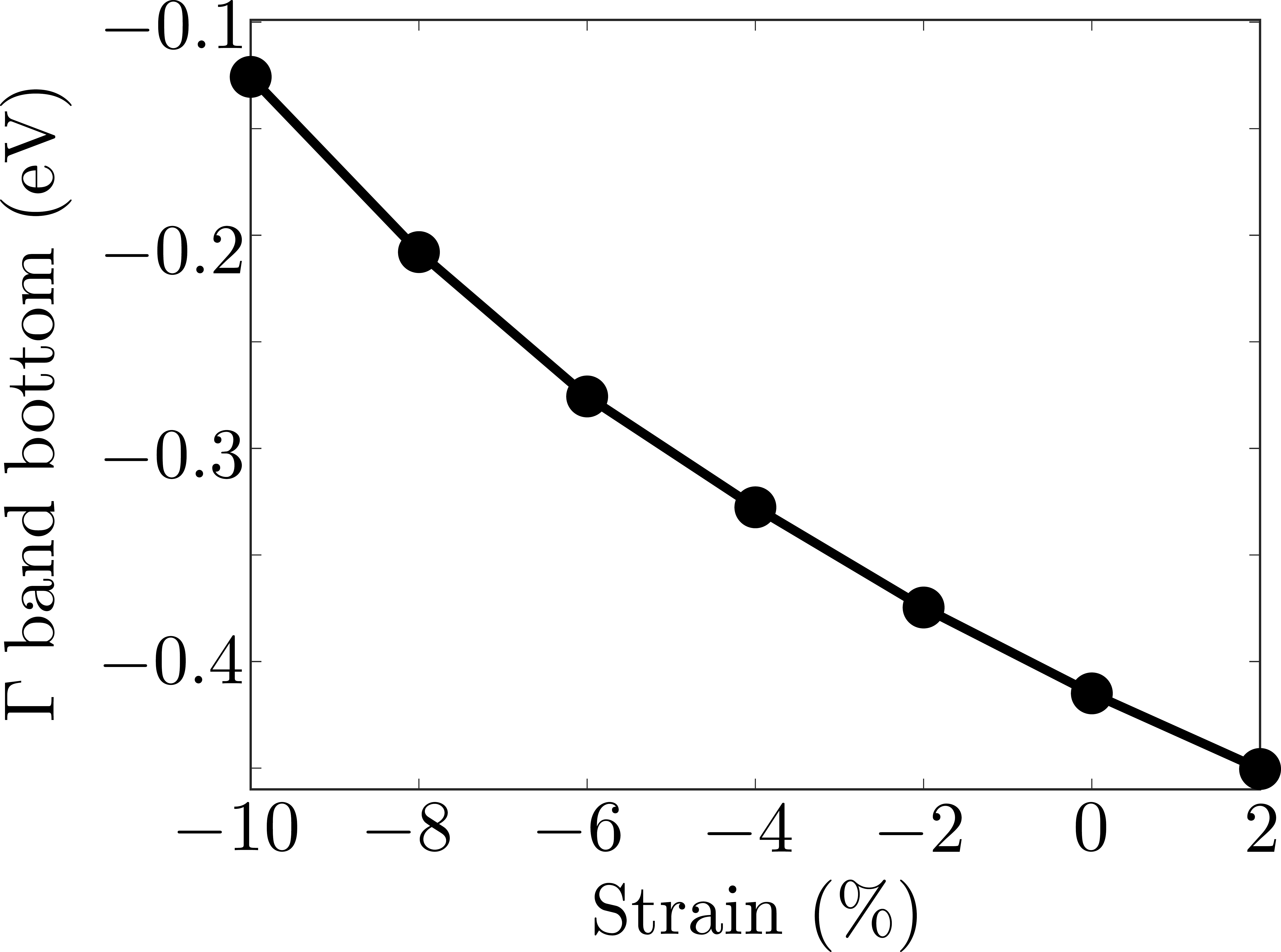}
\caption{\label{fig:gamma_pocket_bottom}DFT results for the shift of the energy of the bottom of the Sb-derived electron band as a function of $c$-axis strain.}
\end{figure}

In addition to the data presented in the main text, in Fig.~\ref{fig:gamma_pocket_bottom} we present the energy of the bottom of the Sb-derived electron band centered at the $\Gamma$ point as a function of the $c$-axis strain. It is clear that strain values of a few percent shift the band by a few hundred meV.

%
%%
%%%
\section{Tight-binding parametrization}
\label{sec:TBmodel}
%%%
%%
%
Our low-energy model consists of a small Sb-derived central electron pocket and a large V-derived hexagonal-like Fermi surface associated with the van Hove singularities (vHs). To describe the latter we employ a simple tight-binding parametrization~\cite{BeugelingEA12}. The electron pocket is assumed to be decoupled from the $d$-orbitals giving rise to the hexagonal band. The effective normal-state Hamiltonian is given by $\pazocal{H} = \sum_{\bo{k}, \sigma} \bo{\Psi}_{\sigma}^{\dagger}(\bo{k}) H_0(\bo{k}) \bo{\Psi}_{\sigma}(\bo{k})$ in the basis where $\bo{\Psi}_{\sigma}^{\dagger}(\bo{k}) = (c_{\bo{k},\mathrm{V}_A \sigma}^{\dagger},~c_{\bo{k},\mathrm{V}_B \sigma}^{\dagger},~c_{\bo{k},\mathrm{V}_C \sigma}^{\dagger},~c_{\bo{k},\mathrm{Sb} \sigma}^{\dagger})$, where $A, B, C$ denote the V sublattice and:
\begin{equation}
    H_0(\bo{k}) = -\begin{bmatrix}
    \mu_{\mathrm{M}} & 2t\cos(k_3) + 2t'\cos(k_1+k_2) & 2t\cos(k_1) + 2t'\cos(k_2+k_3) & 0 \\ 
    2t\cos(k_3) + 2t'\cos(k_1+k_2) & \mu_{\mathrm{M}} & 2t\cos(k_2) + 2t'\cos(k_3-k_1) & 0 \\
    2t\cos(k_1) + 2t'\cos(k_2+k_3) & 2t\cos(k_2) + 2t'\cos(k_3-k_1) & \mu_{\mathrm{M}} & 0 \\
    0 & 0 & 0 & -\xi_{\Gamma}(\bo{k})
    \end{bmatrix}.
    \label{eq:H0}
\end{equation}

\begin{figure}[]
	\centering
	\includegraphics[width=0.3\linewidth]{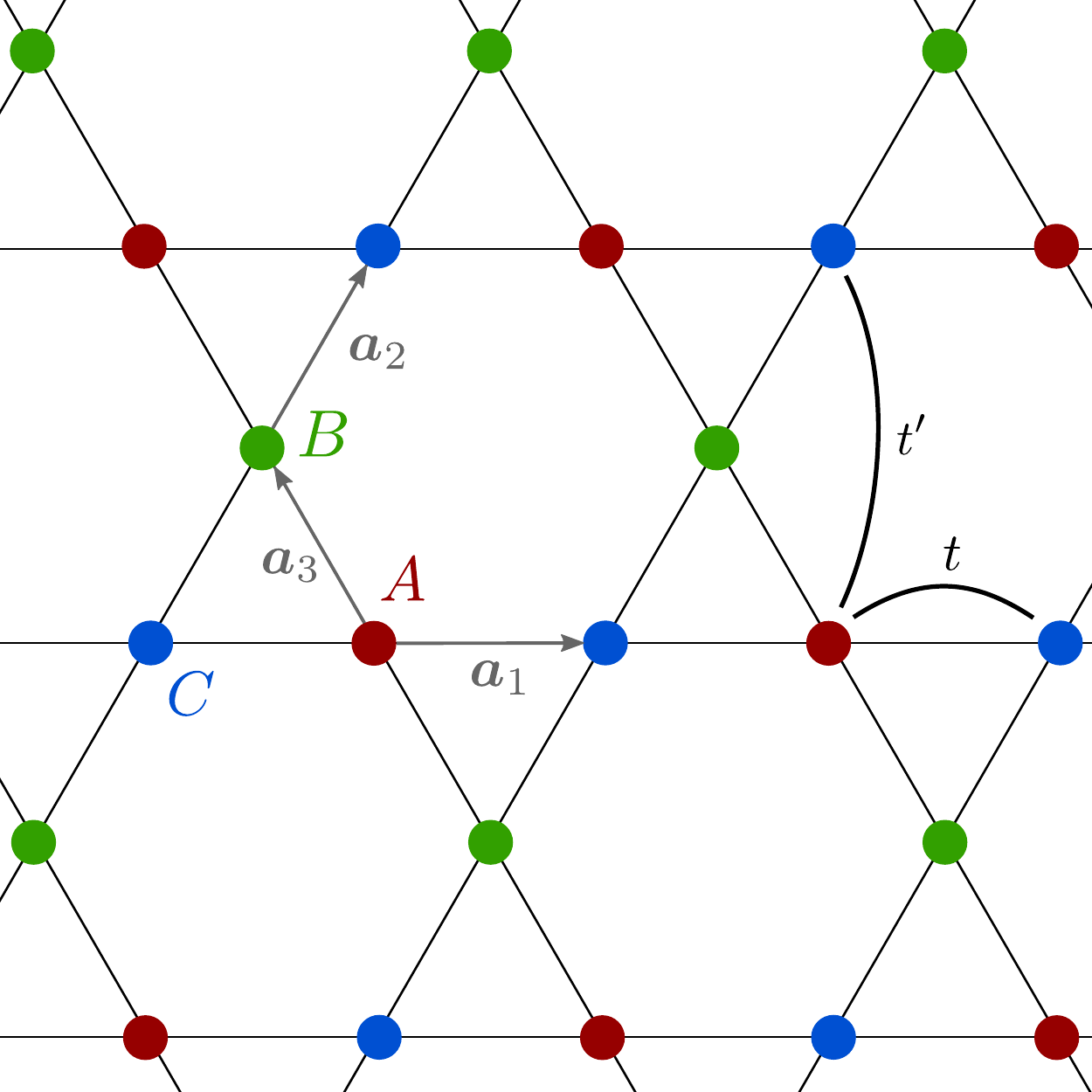}
    \caption{The kagome lattice with the three sublattices ($A, B, C$) indicated. Nearest-neighbor and next-nearest neighbor hoppings ($t, t'$), as well as intra-unit cell vectors ($\bo{a}_1, \bo{a}_2, \bo{a}_3$), are also shown.}
	\label{fig:Kagome}
\end{figure}

Here, $k_i = \bo{k}\cdot \bo{a}_i$, with $\bo{a}_1 = (a/2,~0)$, $\bo{a}_2 = (a/4, \sqrt{3}a/4)$, and $\bo{a}_3 = \bo{a}_2 - \bo{a}_1$, see Fig.~\ref{fig:Kagome}.
The band dispersion from which the $\Gamma$ pocket originates can be conveniently modeled by~\cite{GuoEA09}:
\begin{equation}
    \xi_{\Gamma}(\bo{k}) = -\tilde{t}\Big(1 + \big[4\sum_{j=1}^3 \cos^2{k_j} -3\big]^{1/2} \Big) - \mu_{\Gamma}. \label{eq:xiGamma}
\end{equation} 
Fixing the length- and energy-scales, $a = 1$ and $\tilde{t} = 1$~eV, there are four free model parameters: $\mu_{\mathrm{M}}$, $\mu_{\Gamma}$, $t$, and $t'$. In the $7$-band model based on first principles calculations in Ref.~\onlinecite{GuEA22}, it was found that $t_{\mathrm{NN}}^{\mathrm{V},d_{x^2-y^2}} = -0.4667~$eV (cf.~$t$ above) and $t_{\mathrm{NN}}^{\mathrm{Sb}} = -0.1813~$eV (cf.~$\tilde{t}$ above). We therefore fix the ratio $t/\tilde{t} \approx 2.6$. The other parameters used, $\mu_{\mathrm{M}} = 0.02t$, $t'=-0.2t$, $\mu_{\Gamma} = -3.65\tilde{t}$, were chosen such that the Fermi surface obtained from the tight-binding parametrization in Eq.~\eqref{eq:H0} matches the Fermi surface of CsV$_3$Sb$_5$ measured experimentally in angle-resolved photoemission spectroscopy (ARPES) in Ref.~\onlinecite{KangEA22}.

%
%%
%%%
\section{Patch model and solution of the gap equations}
\label{sec:Symmetries}
\begin{figure}
	\centering
	\includegraphics[width=\linewidth]{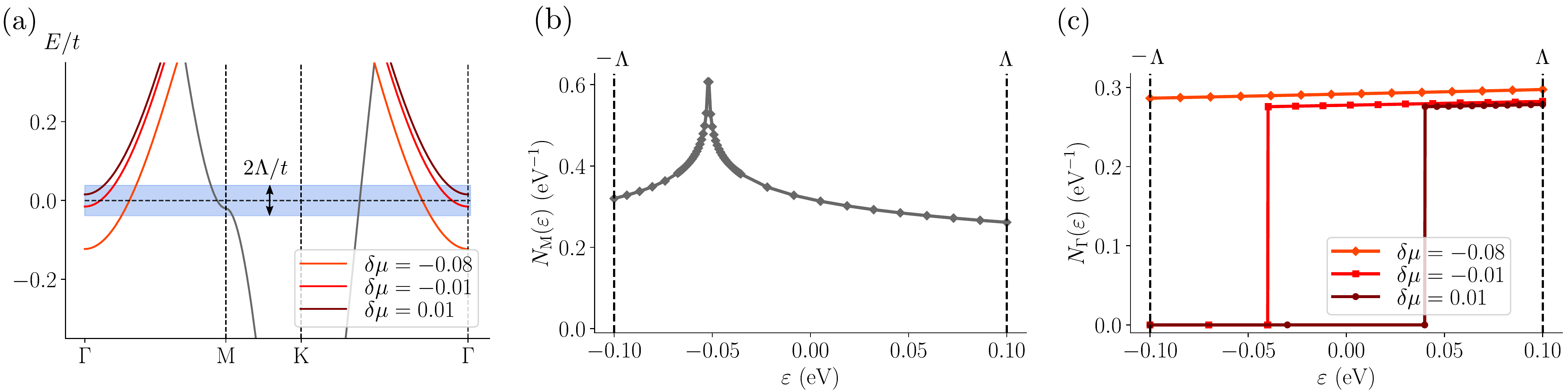}
    \caption{(a) Low-energy band dispersion resulting from the tight-binding parametrization in Eq.~\eqref{eq:H0}. (b) DOS for the hexagonal band, with the (logarithmic and integrable) van Hove singularity shown at $\varepsilon_c = -0.02t$. (c) DOS for the $\Gamma$ pocket at three different values of $\mu_\Gamma$ with corresponding band structures shown in panel (a). The DOS vanishes for $\varepsilon$ smaller than the bottom of the band. In the labels we introduced $\delta \mu \equiv (\mu_{\Gamma} - \mu_c)/\mu_c$, and $\mu_c = -4\tilde{t}$. For the interaction cutoff we used $\Lambda = 100$~meV.} 
	\label{fig:DOS}
\end{figure}

%%%
%%
%
We consider a low-energy patch model where the important degrees of freedom are the states near the three M points, where the vHs are located, and near the $\Gamma$ point, where the small electron pocket is present. Hence, our model consists of four patches, three at M and one at $\Gamma$. The part of the interaction Hamiltonian relevant for superconductivity is given by~\cite{NandkishoreEA12}
\begin{equation}
\begin{aligned}
    H_{\rm int} =& \frac{1}{2}\tilde{g}_{\mathrm{M}\bar{\mathrm{M}}}\sum_{\substack{\alpha\neq\beta, \\ \sigma }} c^{\dagger}_{\mathrm{M}_{\alpha}\sigma}c^{\dagger}_{\mathrm{M}_{\alpha}\bar{\sigma}}c_{\mathrm{M}_{\beta}\bar{\sigma}}^{\vphantom{\dagger}} c_{\mathrm{M}_{\beta}\sigma}^{\vphantom{\dagger}} +    
    \frac{1}{2} \tilde{g}_{\mathrm{MM}} \sum_{\alpha, \sigma} c^{\dagger}_{\mathrm{M}_{\alpha}\sigma}c^{\dagger}_{\mathrm{M}_{\alpha}\bar{\sigma}}c_{\mathrm{M}_{\alpha}\bar{\sigma}}^{\vphantom{\dagger}} c_{\mathrm{M}_{\alpha}\sigma}^{\vphantom{\dagger}}  \\
    & + \frac{1}{2}\tilde{g}_{\Gamma\Gamma}\sum_{\sigma} c^{\dagger}_{\Gamma\sigma}c^{\dagger}_{\Gamma\bar{\sigma}}c_{\Gamma\bar{\sigma}}^{\vphantom{\dagger}} c_{\Gamma\sigma}^{\vphantom{\dagger}}
    + \frac{1}{2}\tilde{g}_{\Gamma \mathrm{M}}\sum_{\alpha, \sigma}\left( c^{\dagger}_{\Gamma\sigma}c^{\dagger}_{\Gamma\bar{\sigma}} c_{\mathrm{M}_{\alpha}\bar{\sigma}}^{\vphantom{\dagger}} c_{\mathrm{M}_{\alpha}\sigma}^{\vphantom{\dagger}} + c^{\dagger}_{\mathrm{M}_{\alpha}\sigma}c^{\dagger}_{\mathrm{M}_{\alpha}\bar{\sigma}}c_{\Gamma\bar{\sigma}}^{\vphantom{\dagger}} c_{\Gamma\sigma}^{\vphantom{\dagger}} \right),
\end{aligned}
\label{eq:Interactions}
\end{equation}
where $\alpha=1,2,3$ runs over the three distinct M points in the Brillouin zone, $\sigma$ denotes spin, and $\bar{\sigma}$ is the opposite of $\sigma$. There are more interactions allowed in the model, but they play no role for superconductivity. Here, $\tilde{g}_{\mathrm{MM}}$, $\tilde{g}_{\mathrm{M} \bar{\mathrm{M}}}$, $\tilde{g}_{\Gamma \mathrm{M}}$, and $\tilde{g}_{\Gamma \Gamma}$ denote dimension-full pairing interactions. As we show below, the only SC instabilities of this model correspond to SC order parameters transforming as the $A_{1g}$ and $E_{2g}$ irreducible representations (irreps) of the point group $D_{6h}$ of the $A$V$_3$Sb$_5$ compounds.
Denoting $\vec{\Delta} \equiv (\Delta_{\rm M_1}, \Delta_{\rm M_2}, \Delta_{\rm M_3}, \Delta_{\Gamma}
)^{\mathrm{T}}$ as the gap functions on the four patches, the corresponding linearized gap equations can be written in matrix form as $\chi_{\mathrm{pp}} \vec{\Delta} = \vec{\Delta}$, with
\begin{equation}
\chi_{\mathrm{pp}} = - \begin{bmatrix}
V^{\mathrm{M}} \tilde{g}_{\mathrm{MM}} & V^{\mathrm{M}} \tilde{g}_{\mathrm{M} \bar{\mathrm{M}}} & V^{\mathrm{M}} \tilde{g}_{\mathrm{M} \bar{\mathrm{M}}} & V^{\Gamma} \tilde{g}_{\Gamma \mathrm{M}} \\
V^{\mathrm{M}} \tilde{g}_{\mathrm{M} \bar{\mathrm{M}}} & V^{\mathrm{M}} \tilde{g}_{\mathrm{MM}} & V^{\mathrm{M}} \tilde{g}_{\mathrm{M} \bar{\mathrm{M}}} & V^{\Gamma} \tilde{g}_{\Gamma \mathrm{M}} \\
V^{\mathrm{M}} \tilde{g}_{\mathrm{M} \bar{\mathrm{M}}} & V^{\mathrm{M}} \tilde{g}_{\mathrm{M} \bar{\mathrm{M}}}& V^{\mathrm{M}} \tilde{g}_{\mathrm{MM}} & V^{\Gamma} \tilde{g}_{\Gamma \mathrm{M}} \\
V^{\mathrm{M}} \tilde{g}_{\Gamma \mathrm{M}} &  V^{\mathrm{M}} \tilde{g}_{\Gamma \mathrm{M}} & V^{\mathrm{M}} \tilde{g}_{\Gamma \mathrm{M}} &  V^{\Gamma} \tilde{g}_{\Gamma \Gamma} 
\end{bmatrix}, \label{eq:GmatrixProper}
\end{equation}
where $V^{\mathrm{M}} \equiv \int_{-\Lambda}^{\Lambda} \D\varepsilon~N_{\mathrm{M}}(\varepsilon) \frac{ \tanh(\beta \varepsilon/2)}{2\varepsilon}$, and $V^{\Gamma} \equiv \int_{ - \Lambda}^{\Lambda} \D\varepsilon~N_{\Gamma}(\varepsilon)  \frac{\tanh(\beta \varepsilon/2)}{2\varepsilon}$. Here, $N_i(\varepsilon)$ are the DOS of the patches, calculated from the tight-binding parametrization as:
\begin{equation}
N_i(\varepsilon) = \frac{\sqrt{3}}{2(2\pi)^2} \int_{\mathrm{BZ}} \D\bo{k}~\delta \left( \varepsilon -  \xi_{i}(\bo{k}) \right) = \frac{\sqrt{3}}{2(2\pi)^2} \int_{S_i(\varepsilon)} \D\bo{k}~1/v_{i}(\bo{k}),
\label{eq:DOS}
\end{equation}
where $v_{i}(\bo{k}) = \lvert \nabla \xi_{i}(\bo{k}) \rvert$ is the Fermi velocity of band $i$, and $S_i(\varepsilon)$ is the constant energy contour $\xi_i(\bo{k})=\varepsilon$. For the $\Gamma$ pocket, $\xi_{\Gamma}$ is given by Eq.~\eqref{eq:xiGamma} whereas for the M pocket, it is given by the second largest eigenvalue of the $3\times3$ upper block-diagonal of the matrix in Eq.~\eqref{eq:H0}, corresponding to the band crossing the Fermi level, as shown in Fig.~\ref{fig:DOS}(a).

In Fig.~\ref{fig:DOS}(b)-(c) we show the DOS for the two bands, as evaluated from Eq.~\eqref{eq:DOS} when integrating over $1/6$th of the energy contours $S_i(\varepsilon)$ (obtained using Brent's method) represented by $400$ $\bo{k}$ points. Panel (c) shows how the DOS of the $\Gamma$ pocked vanishes as the bottom of the electron pocket is tuned through the incipient band regime by increasing the parameter $\lvert \mu_\Gamma \rvert$, which mimics the effect of hole doping. Note that, according to our convention, the Fermi level is always at $\varepsilon=0$.
\begin{figure}
	\centering
    \includegraphics[width=\linewidth]{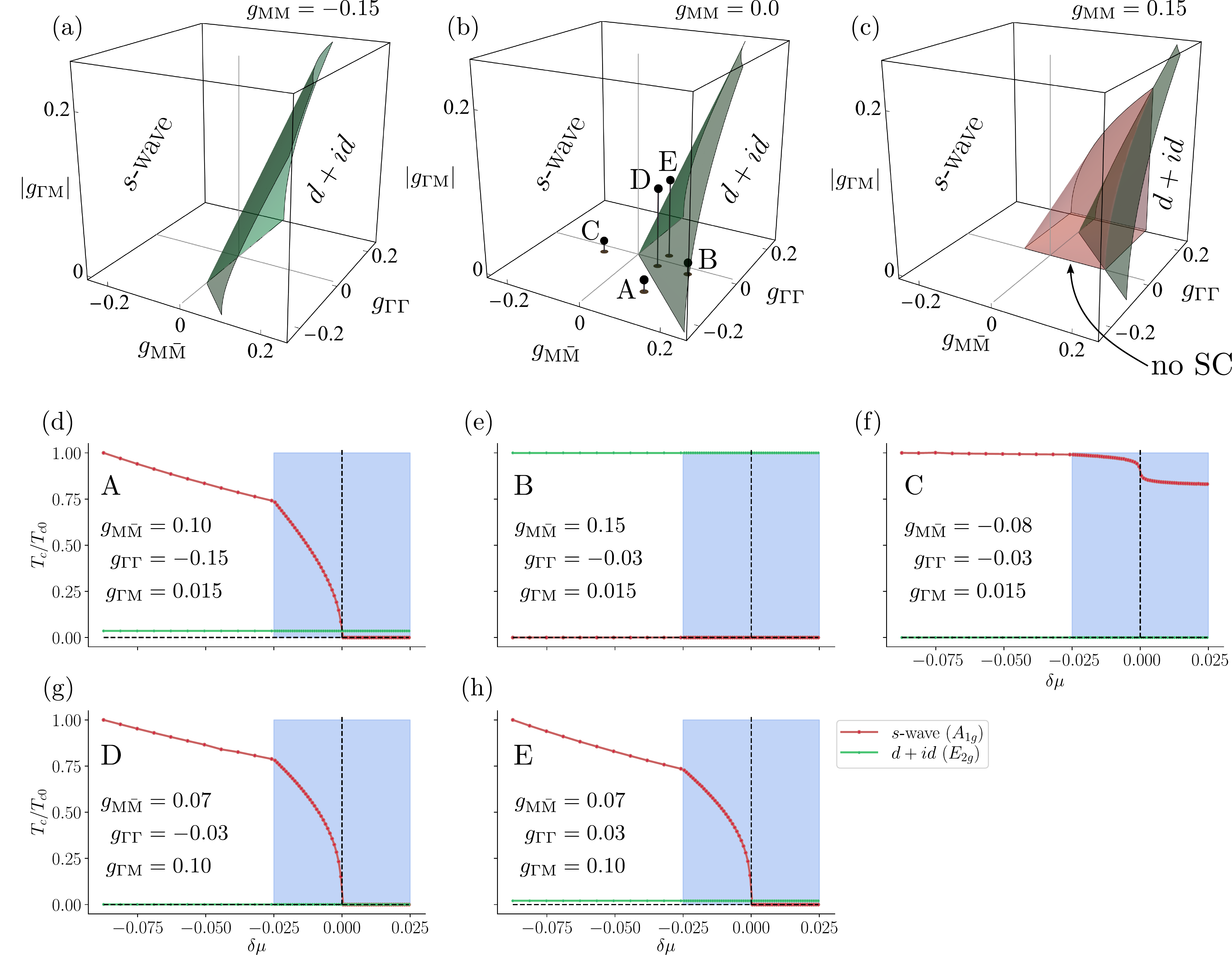}
    \caption{Panels (a)-(c): Phase diagrams obtained from Eq.~\eqref{eq:Gmatrix} for different values of $g_{\rm MM}$. Panel (b) corresponds to the figure shown in the main text. Panels (d)-(h): Critical temperature $T_c$ for the two leading SC channels for interaction parameters corresponding to the symbols in the phase diagram of panel (b). The critical temperatures were obtained by setting the maximal eigenvalue of Eq.~\eqref{eq:GmatrixProper} to be $1$, and were normalized by the value $T_{c0}$ obtained at $\mu_{\Gamma 0} = -3.65 \Tilde{t}$. Note that $T_{c0}$ is different for each set of $g_i$ values. Panels (d) and (g) correspond to the cases shown in the main text. Here, $\delta \mu \equiv (\mu_{\Gamma} - \mu_c)/\mu_c$, and $\mu_c = -4\tilde{t}$.}
	\label{fig:GapRatio}
\end{figure}

Except when very close to the Lifshitz transition characterized by the vanishing of the $\Gamma$ pocket, it is safe to approximate $N_i(\varepsilon)\approx N_i(0)$. Then, $V^{\mathrm{M}} = N_{\mathrm{M}}(0) V_{\Lambda}$ and $V^{\Gamma} = N_\Gamma(0) V_{\tilde{\mu}}$, with $\tilde{\mu} \equiv \min \left( \mu_\Gamma - \mu_c, \Lambda \right)$ and $V_a$ given by:

\begin{equation}
    V_{a} = \int_{-a}^{\Lambda} \D\varepsilon~\f{\tanh(\beta \varepsilon/2)}{2\varepsilon} \approx
    \begin{cases}
     \ln\left(c_0 \beta \sqrt{a \Lambda} \right) & \mathrm{if}~0< a \leq \Lambda, \\
     \ln\left(\Lambda/\lvert a \rvert \right) & \mathrm{if}~a < 0,
    \end{cases}
\label{eq:AsymptoticBCS}
\end{equation}
Here, asymptotic expressions were derived in the last step, where we defined $c_0 \equiv 2\exp(\gamma)/\pi \approx 1.13$ with $\gamma$ denoting Euler's constant. Notably, the logarithmic divergence disappears in the ``deep incipient band'' regime, corresponding to $\mu_\Gamma < \mu_c$ (see also Ref.~\onlinecite{ChenEA15}). Exploiting the asymptotic behavior in Eq.~\eqref{eq:AsymptoticBCS} in the ``deep incipient band regime'' one can derive a weak-coupling expression for the chemical potential at which the critical temperature of the $s$-wave solution drops exponentially to zero: 
\begin{equation}
\mu_{\Gamma,\mathrm{crit}} \approx \mu_c - \Lambda \exp(-[g_{\mathrm{MM}}+2g_{\mathrm{M\bar{M}}}] / [3g_{\Gamma \mathrm{M}}^2] ),
\label{eq:muCrit}
\end{equation}
which is qualitatively similar to the expression found in Ref.~\onlinecite{ChenEA15}.

Focusing on the non-incipient-band case, where $\mu_{\Gamma}-\mu_c > \Lambda$ and the BCS regime is appropriate ($\beta \Lambda \gg 1$), the gap equations simplify considerably, since $V_{\tilde{\mu}} = V_\Lambda$. Defining the dimensionless pairing interactions $g_{\mathrm{MM}} = \tilde{g}_{\mathrm{MM}} N_{\mathrm{M}}(0)$, $g_{\mathrm{M}\bar{\mathrm{M}}} = \tilde{g}_{\mathrm{M}\bar{\mathrm{M}}} N_{\mathrm{M}}(0)$, $g_{\Gamma \Gamma} = \tilde{g}_{\Gamma \Gamma} N_\Gamma(0)$ and $g_{\Gamma \mathrm{M}} = \tilde{g}_{\Gamma \mathrm{M}} \sqrt{N_{\mathrm{M}}(0) N_\Gamma(0)}$, we obtain the gap equations of the main text, with $\eta \equiv \sqrt{N_\Gamma(0)/N_{\mathrm{M}}(0)}$ and:
\begin{equation}
\chi_{\mathrm{pp}} = - V_{\Lambda} \begin{bmatrix}
 g_{\mathrm{MM}} & g_{\mathrm{M} \bar{\mathrm{M}}} &  g_{\mathrm{M} \bar{\mathrm{M}}} &  \eta g_{\Gamma \mathrm{M}} \\
 g_{\mathrm{M} \bar{\mathrm{M}}} & g_{\mathrm{MM}} &  g_{\mathrm{M} \bar{\mathrm{M}}} & \eta g_{\Gamma \mathrm{M}} \\
 g_{\mathrm{M} \bar{\mathrm{M}}} & g_{\mathrm{M} \bar{\mathrm{M}}}& g_{\mathrm{MM}} & \eta g_{\Gamma \mathrm{M}} \\
 \eta^{-1} g_{\Gamma \mathrm{M}} &  \eta^{-1} g_{\Gamma \mathrm{M}} &  \eta^{-1} g_{\Gamma \mathrm{M}} & g_{\Gamma \Gamma} 
\end{bmatrix}\,. \label{eq:Gmatrix}
\end{equation}
Diagonalization of $\chi_{\mathrm{pp}} / V_\Lambda$ yields the two leading and competing eigenvalues:
\begin{equation}
\begin{aligned}
    \lambda_{1} & = g_{\mathrm{M} \bar{\mathrm{M}}} - g_{\mathrm{MM}}, \\
    \lambda_{2} &= \frac{1}{2} \left(\tilde{g} -g_{\Gamma \Gamma} - g_{\mathrm{MM}} - 2g_{\mathrm{M} \bar{\mathrm{M}}} \right),
\end{aligned}
\label{eq:Eigenvalues}
\end{equation}
where
\begin{equation}
    \tilde{g} =
    \sqrt{12 g_{\Gamma M}^2 + \left( g_{\Gamma \Gamma} - g_{\mathrm{MM}} - 2 g_{\mathrm{M} \bar{\mathrm{M}}} \right)^2 }.
\end{equation}
We note that the eigenvalue $\lambda_1$ is doubly degenerate, and that there is a fourth eigenvalue $\lambda_3 = \lambda_2 - \tilde{g}$ which we do not consider, as it is always smaller than $\lambda_2$. For each eigenvalue $\lambda_{i}$, as long as it is positive, there is a superconducting instability with critical temperature $T_c^{i}  = \frac{2 \mathrm{e}^\gamma}{k_B \pi} \Lambda \exp\left( -1/\lambda_{i}\right)$. To determine the SC channels corresponding to these eigenvalues, we compute the corresponding eigenvectors (not normalized), which are given by
\begin{equation}
    \vec{\Delta}_{1}^{(1)} = 
    \begin{bmatrix}
        -1 \\ 1 \\ 0 \\ 0
    \end{bmatrix}, \hspace{10pt} 
    \vec{\Delta}_{1}^{(2)} =
    \begin{bmatrix}
        -1 \\ -1 \\ 2 \\ 0
    \end{bmatrix},
     \hspace{10pt}
    \vec{\Delta}_{2} =
    \begin{bmatrix}
    1  \\ 1 \\ 1 \\ -\sqrt{3} \, \mathrm{sign}\left(g_{\Gamma \mathrm{M}}\right) \eta^{-1} \zeta
    \end{bmatrix}, 
\label{eq:Eigenvectors}
\end{equation}
where
\begin{equation}
\zeta \equiv \sqrt{1+\left( \frac{g_{\Gamma \Gamma} - g_{\mathrm{MM}} - 2 g_{\mathrm{M} \bar{\mathrm{M}}}}{\sqrt{12} g_{\Gamma \mathrm{M}}} \right)^2} - 
\left( \frac{g_{\Gamma \Gamma} - g_{\mathrm{MM}} - 2 g_{\mathrm{M} \bar{\mathrm{M}}}}{\sqrt{12} \lvert g_{\Gamma \mathrm{M}} \rvert} \right)
 > 0.
\label{eq:Zeta}
\end{equation}
The first two eigenvectors correspond to the $d_{xy}$ ($\vec{\Delta}_{1}^{(1)}$) and $d_{x^2-y^2}$ ($\vec{\Delta}_{2}^{(2)}$) form factors. As such, they transform  as the two-dimensional irreducible representation $E_{2g}$, and the degenerate eigenvalue $\lambda_1$ is associated with a $d$-wave SC instability~\cite{NandkishoreEA12}. As for the eigenvector $\vec{\Delta}_{2}$, it transforms as the one-dimensional trivial irreducible presentation $A_{1g}$, implying that the eigenvalue $\lambda_2$ is associated with an $s$-wave SC instability. Importantly, the structure of this $s$-wave state depends on the sign of the pairing interaction $g_{\Gamma \mathrm{M}}$. As shown in Eq.~\eqref{eq:Eigenvectors}, $\Delta_\Gamma$ has an opposite (equal) sign as $\Delta_{\mathrm{M}}$ for $g_{\Gamma \mathrm{M}} > 0$ ($g_{\Gamma \mathrm{M}} < 0$), corresponding to an $s^{+-}$ ($s^{++}$) pairing state.

The relative phase and relative amplitude between the degenerate $d_{xy}$ and $d_{x^2-y^2}$ gap functions can only be found by going beyond the linearized gap equations. In practice, one computes the coefficients of the biquadratic couplings between the two gaps in the Landau free energy expansion. The analysis in the subspace of the $\mathrm{M}$ patches was done in Ref.~\cite{NandkishoreEA12}, which found a relative phase of $\pi/2$ between the two gap functions, implying a time-reversal symmetry-breaking $d_{xy} \pm id_{x^2-y^2}$ state. Microscopic evaluations of the quartic order Ginzburg--Landau coefficients also find under quite generic circumstances the time-reversal breaking state to be favored~\cite{Sauls94}.

The phase diagram shown in the main text, and repeated in Fig. \ref{fig:GapRatio}(b), corresponds to the regime away from the Lifshitz transition of the $\Gamma$ pocket, where Eq.~\eqref{eq:Gmatrix} holds. Figs.~\ref{fig:GapRatio}(a) and (c) show the impact of the intra-M-patch interaction on the phase diagram: an attractive $g_{\mathrm{M} \mathrm{M}}$ enhances the range of $d+id$ state, whereas a repulsive $g_{\mathrm{M} \mathrm{M}}$ not only suppresses the $d+id$ state, but also completely kills superconductivity in parts  the phase diagram. 

The evolution of $T_c$ with increasing $\delta \mu \equiv (\mu_{\Gamma} - \mu_c)/\mu_c$ for different points indicated in the phase diagram of Fig.~\ref{fig:GapRatio}(b) are shown in Figs.~\ref{fig:GapRatio}(d)-(h). For each dimensionless parameter $g_i$, the dimensionfull parameter $\tilde{g}_i$, which enters the gap equation~\eqref{eq:GmatrixProper}, was calculated according to the definition from the density of states at the Fermi level calculated for the $\Gamma$ and $\mathrm{M}$ bands for $\mu_{\Gamma 0} = -3.65 \tilde{t}$. It is clear that the $s$-wave state promoted by a dominant attractive $g_{\mathrm{M} \bar{\mathrm{M}}}$ is robust across the Lifshitz transition, since pairing is little affected by the $\Gamma$ pocket, see Fig.~\ref{fig:GapRatio}(f). Similarly, the $d+id$ state is essentially unaffected by the Lifshitz transition, since it is promoted by a dominant repulsive $g_{\mathrm{M} \mathrm{M}}$ [Fig.~\ref{fig:GapRatio}(e)]. 

There are only two scenarios in which pairing is strongly suppressed across the Lifshitz transition: a dominant attractive $g_{\Gamma \Gamma}$ or a dominant inter-patch $g_{\Gamma \mathrm{M}}$, be it repulsive or attractive. These scenarios are shown in Figs.~\ref{fig:GapRatio}(d) and (g)-(h), respectively. The last two panels show that the sign of the intra-$\Gamma$-patch interaction does not significantly impact the evolution of $T_c$ when $g_{\Gamma \mathrm{M}}$ is the dominant interaction. Note that, in Figs.~\ref{fig:GapRatio}(d) and (g)-(h), a large $g_{\mathrm{M} \mathrm{M}}$ would imply a transition from an $s$-wave to a $d+id$-wave state upon crossing the Lifshitz transition, which is not observed experimentally.  

\end{document}